\documentclass{article}

\usepackage{mathrsfs}
\usepackage{amsfonts}
\usepackage{amsmath}
\usepackage{amssymb}
\usepackage{graphicx}

\title{How to create a universe}
\author{Gordon McCabe}

\begin{document}

\maketitle

\begin{abstract}

The purpose of this paper is (i) to expound the specification of a
universe, according to those parts of mathematical physics which have
been experimentally and observationally verified in our own universe;
and (ii) to expound the possible means of creating a universe in the
laboratory.

\end{abstract}

\section{Universe specification}

According to modern mathematical theoretical physics, the
specification of a physical universe can be broken down into the
following:

\begin{enumerate}
\item{Specify a space-time.}
\item{Specify a set of gauge fields.}
\item{Specify a set of elementary particles, and partition them into a finite number of generations.}
\item{Specify the strengths of the gauge fields.}
\item{Specify the couplings between the gauge fields and the elementary particles.}
\item{Specify the direct (`Yukawa') couplings between elementary particles.}
\item{Specify the mixing between the elementary particles in different generations.}
\item{Specify cosmological parameters corresponding to the initial conditions for the universe.}

\end{enumerate}

This list is based upon the understanding gleaned from general
relativistic cosmology and the standard model of particle physics,
the latter being an application of quantum field theory. Both
these theories are empirically verified. I do not intend to
consider how one might define a physical universe according to
speculative theories such as string theory or supersymmetry. Note
also that these specifications are not necessarily independent;
for example, the parameters required to uniquely specify an
elementary particle include those which specify its coupling to
the gravitational field and the gauge fields (mass, electric
charge, weak hypercharge etc.), and parameters which depend upon
the dimension and signature of space-time (spin, parity etc.).

Each one of the specifications above involves making a choice from a
huge range of possibilities on offer. As a consequence, there exists
a huge set of possible physical universes. In this paper, we will
approach the notion of possible physical universes using the
philosophical doctrine of `structural realism', which asserts that,
in mathematical physics at least, the physical domain of a true
theory is an instance of a mathematical structure. It follows that if
the domain of a true theory extends to the entire physical universe,
then the entire universe is an instance of a mathematical structure.
Equivalently, it is asserted that the physical universe is isomorphic
to a mathematical structure.

In terms of the specifications above, the mathematical structures
are as follows:

\begin{enumerate}
\item{To specify a space-time, a pseudo-Riemannian manifold $(\mathcal{M},g)$ of a particular dimension $n$ and signature $(p,q)$ is specified.}
\item{To specify a gauge field, a compact connected Lie group $G$ is specified. This is called the gauge group of the field.}
\item{To specify a set of elementary particles, and their partition into a finite number of generations, a
finite family of irreducible unitary representations of the local
space-time symmetry group $\mathscr{P}$ is specified.}
\item{To specify the strength of a gauge field with gauge group $G$, a choice of adjoint-invariant metric is specified in the Lie
algebra $\mathfrak{g}$.}
\item{To specify the couplings between the gauge fields and the elementary particles, the values of
the `charges' possessed by elementary particles are specified by
means of finite-dimensional irreducible representations of the
gauge groups of the corresponding gauge fields.}
\item{To specify the direct (`Yukawa') couplings between elementary particles, Yukawa matrices are specified.}
\item{To specify the mixing between the elementary particles in different generations, matrices such
as the Cabibbo-Kobayashi-Maskawa matrix are specified.}
\item{To specify cosmological parameters corresponding to the initial conditions for the universe,
the space-time $(\mathcal{M},g)$ is taken to be a product manifold
$\mathbb{R} \times \Sigma$, the global symmetry group (or Killing Lie
algebra) of $3$-dimensional space $\Sigma$ is specified, and the
dynamical parameters which determine the time evolution of the
3-dimensional geometry, such as the Hubble parameter $H_0$ and
density parameter $\Omega_0$, are specified.}

\end{enumerate}

Our own physical universe is specified as follows:

\begin{enumerate}
\item{Space-time is a $4$-dimensional pseudo-Riemannian manifold $(\mathcal{M},g)$ of signature $(3,1)$.}
\item{There are three force fields: the strong nuclear force, the weak nuclear force, and the electromagnetic
force. The weak and electromagnetic fields are unified in the
electroweak gauge field. The gauge group of the strong force is
$SU(3)$; the gauge group of the electroweak force is $U(2) \cong
SU(2) \times U(1)/\mathbb{Z}_2$, and the gauge group of the
electromagnetic force is $U(1)$.}
\item{The elementary particles consist of quarks, leptons and gauge
bosons. Elementary particles are divided into fermions and bosons
according to the value they possess of a property called
`intrinsic spin'. If a particle possesses a non-integral value of
intrinsic spin, it is referred to as a fermion, whilst if it
possesses an integral value, it is referred to as a boson. The
particles of the elementary matter fields are fermions and the
interaction carriers of the gauge force fields are bosons. The
elementary fermions come in two types: leptons and quarks. Whilst
quarks interact via both the strong and electroweak forces,
leptons interact via the electroweak force only. There are six
types of lepton and six types of quark. The six leptons consist of
the electron and electron-neutrino $(e,\nu_e)$, the muon and
muon-neutrino $(\mu, \nu_{\mu})$, and the tauon and tauon-neutrino
$(\tau,\nu_{\tau})$. The six quarks consist of the up-quark and
down-quark $(u,d)$, the charm-quark and strange-quark $(c,s)$, and
the top-quark and bottom-quark $(t,b)$. The six leptons have six
anti-leptons, $(e^+,\overline{\nu}_e)$,
$(\mu^+,\overline{\nu}_{\mu})$, $(\tau^+,\overline{\nu}_{\tau})$,
and the six quarks have six anti-quarks
$(\overline{u},\overline{d})$, $(\overline{c},\overline{s})$,
$(\overline{t},\overline{b})$. The masses of the elementary
fermions are as follows:\footnote{All the parameter values and
estimates in this paper are taken from Tegmark \textit{et al}
2005.}

\begin{center}
{
\begin{tabular}{|l|ll|}
\hline
$m_e            $   &Electron mass                                     &$(510998.92\pm 0.04)\text{eV}$\\
$m_\mu          $   &Muon mass                                   &$(105658369\pm 9)\text{eV}$\\
$m_\tau         $   &Tauon mass                                  &$(1776.99\pm 0.29)\text{MeV}$\\
\cline{2-3}
$m_u            $   &Up quark mass                                    &$(1.5-4)\text{MeV}$\\
$m_d            $   &Down quark mass                                    &$(4-8)\text{MeV}$\\
$m_c            $   &Charm quark mass                                   &$(1.15-1.35)\text{GeV}$\\
$m_s            $   &Strange quark mass                                 &$(80-130)\text{MeV}$\\
$m_t            $   &Top quark mass                                     &$(174.3\pm 5.1)\text{GeV}$\\
$m_b            $   &Bottom quark mass                                 &$(4.1-4.9)\text{GeV}$\\
\cline{2-3}
$m_{\nu_e}  $      &Electron neutrino mass                               &$<3 \text{eV}$\\
$m_{\nu_\mu}    $      &Muon neutrino mass                            &$<0.19 \text{MeV}$\\
$m_{\nu_\tau}   $      &Tau neutrino mass                           &$<18.2 \text{GeV}$\\
\hline
\end{tabular}
} % \tiny
\end{center}
}

\item{To specify the strength of a gauge field with gauge group $G$,
a choice of adjoint-invariant metric is specified in the Lie algebra
$\mathfrak{g}$. The degrees of freedom available in the choice of
this metric correspond to what physicists call the `coupling
constants' of the gauge field. In the case of a gauge field with a
simple gauge group, there is a single degree of freedom in the choice
of the adjoint-invariant metric upon the corresponding Lie algebra,
hence there is a single coupling constant. For a gauge field with a
more general compact gauge group $G$, there is a coupling constant
for every simple Lie algebra and every copy of $\mathfrak{u}(1)$ in a
direct sum decomposition of the Lie algebra $\mathfrak{g}$. The gauge
group of the electromagnetic field is $U(1)$, hence there is a single
electromagnetic coupling constant, determined by $q$, the charge of
the electron. The gauge group of the strong force, $SU(3)$, is
simple, hence the strong force also has a single coupling constant,
$g_s$. In the case of the electroweak force, with gauge group $U(2)
\cong SU(2)_L \times U(1)_Y/\mathbb{Z}_2$, there are two coupling
constants: $g$, the weak isospin coupling constant, associated with
$SU(2)_L$, and $g'$, the weak hypercharge coupling constant,
associated with $U(1)_Y$. Alternatively, one can specify the metric
on $\mathfrak{u}(2)$ with a combination of the Weinberg angle
$\theta_W$, and the charge of the electron $q$. These parameters are
related by the expressions $g = q/\sin \theta_W$ and $g' = q/\cos
\theta_W$.

\begin{center}
{
\begin{tabular}{|l|ll|}
\hline
$g      $   &Weak isospin coupling constant at $m_Z$     &$0.6520\pm 0.0001$\\
$\theta_W   $   &Weinberg angle              &$0.48290\pm 0.00005$\\
$g_s        $   &Strong coupling constant at $m_Z$   &$1.221\pm
0.022$\\ \hline
\end{tabular}
} % \tiny
\end{center}
}

\item{To specify the couplings between the gauge fields and the elementary particles, the values of
the `charges' possessed by elementary particles are specified. For
example, in the case of couplings between particles and the
electromagnetic force, the strength of the coupling is determined
by the electromagnetic charge of the particle. In general, the
charges of an elementary particle correspond to the irreducible
representation of $SU(3) \times SU(2) \times U(1)$ with which that
particle is associated. Particles which lack a certain type of
charge will not couple at all to the corresponding gauge field.}

\item{In the case of a universe with spontaneous symmetry breaking
caused by Higgs fields, the Yukawa couplings specify the direct
interactions between the Higgs bosons and the elementary fermions.
The nature of these interactions are specified by trilinear invariant
forms upon the typical fibre of tensor product bundles such as $\iota
\sigma_L \otimes \iota \otimes (\Lambda^2
\overline{\iota})\overline{\sigma}_R$ (Derdzinski p188), where
$\sigma_L$ is a left-handed Weyl spinor bundle, $\sigma_R$ is a
right-handed Weyl spinor bundle, and $\iota$ is an electroweak
interaction bundle, a complex plane bundle. The coefficients which
specify these trilinear forms are the Yukawa couplings, and these
couplings are often organised into Yukawa matrices. The masses of the
elementary fermions are related to the Yukawa coupling coefficients,
listed as follows:

\begin{center}
{
\begin{tabular}{|l|ll|}
\hline
$G_e            $   &Electron Yukawa coupling        &$2.94\times 10^{-6}$\\
$G_\mu          $   &Muon Yukawa coupling            &$0.000607$\\
$G_\tau         $   &Tauon Yukawa coupling           &$0.0102156233$\\
\cline{2-3}
$G_u            $   &Up quark Yukawa coupling        &$0.000016\pm 0.000007$\\
$G_d            $   &Down quark Yukawa coupling      &$0.00003\pm 0.00002$\\
$G_c            $   &Charm quark Yukawa coupling         &$0.0072\pm 0.0006$\\
$G_s            $   &Strange quark Yukawa coupling       &$0.0006\pm 0.0002$\\
$G_t            $   &Top quark Yukawa coupling       &$1.002\pm 0.029$\\
$G_b            $   &Bottom quark Yukawa coupling        &$0.026\pm 0.003$\\
\cline{2-3}
$G_{\nu_e}  $      &Electron neutrino Yukawa coupling &$<1.7\times 10^{-11}$\\
$G_{\nu_\mu}    $      &Muon neutrino Yukawa coupling &$<1.1\times10^{-6}$\\
$G_{\nu_\tau}   $      &Tau neutrino Yukawa coupling &$<0.10$\\
\hline
\end{tabular}
} % \tiny
\end{center}

The electroweak Higgs field $\phi$ itself requires two parameters,
$\lambda$ and $\mu$, for its specification, and these parameters
determine the masses of the Higgs bosons and the electroweak gauge
bosons.

\begin{center}
{
\begin{tabular}{|l|ll|}
\hline $\mu^2      $   &Quadratic Higgs coefficient         &$\sim
-10^{-33}$\\
$\lambda    $   &Quartic Higgs coefficient       &$\sim 1$?\\
\hline
\end{tabular}
} % \tiny
\end{center}

}
\item{The mixing between the quarks in different generations is specified by the Cabibbo-Kobayashi-Maskawa (CKM) matrix.
By convention, the mixing is expressed in terms of the $\{d,s,b\}$
quark flavours. The bundle which represents the generalisation of
these three quark flavours is $\sigma_d \oplus \sigma_s \oplus
\sigma_b$, where each summand is a copy of the Dirac spinor bundle
$\sigma$. Different mixtures of these flavours correspond to
different orthogonal decompositions $\sigma_{d'} \oplus \sigma_{s'}
\oplus \sigma_{b'}$. Each such decomposition is defined by the
Cabibbo-Kobayashi-Maskawa matrix, which can be specified by four
parameters, $\{ \theta_{12}, \theta_{23}, \theta_{13}, u \}$, called
the Cabibbo-Kobayashi-Maskawa parameters. The first three parameters
$\theta_{12}, \theta_{23}, \theta_{13}$ are angular parameters with
values in $[0,\frac{\pi}{2}]$. The fourth parameter is a phase factor
$u = e^{i \delta}$, (Derdzinski 1992, p160). This notion of quark
mixing is considered to be a consequence of the interaction of quarks
with the electroweak Higgs bosons. If, as current evidence indicates,
the neutrinos possess mass, then there is a corresponding notion of
lepton mixing, and the CKM matrix has a lepton counterpart called the
Maki-Nakagawa-Sakata (MNS) matrix. This matrix also requires four
parameters for its specification.

\begin{center}
{
\begin{tabular}{|l|ll|}
\hline
$\sin\theta_{12}$   &Quark CKM matrix angle          &$0.2243\pm 0.0016$\\
$\sin\theta_{23}$   &Quark CKM matrix angle          &$0.0413\pm 0.0015$\\
$\sin\theta_{13}$   &Quark CKM matrix angle          &$0.0037\pm 0.0005$\\
$\delta_{13}    $   &Quark CKM matrix phase          &$1.05\pm 0.24$\\
\cline{2-3}
$\sin\theta_{12}'$  &Neutrino MNS matrix angle       &$0.55\pm 0.06$\\
$\sin\theta_{23}'$    &Neutrino MNS matrix angle       &$\ge 0.94$\\
$\sin\theta_{13}'$  &Neutrino MNS matrix angle       &$\le 0.22$\\
$\delta_{13}'   $   &Neutrino MNS matrix phase       &$?$\\
\hline
\end{tabular}
}
\end{center}
}
\item{3-dimensional space on cosmological scales is currently thought to be well-approximated
by $\mathbb{R}^3$, with $H_0 \approx (10 \; \text{Gyr})^{-1}$ and
$\Omega_0 = 1.01 \pm 0.02$.}

\end{enumerate}

\section{Universe creation in a laboratory}

Inflationary cosmology postulates that there was a period in our
universe's early history during which gravitation became effectively
repulsive, and the universe consequently underwent exponential
expansion (see Blau and Guth, 1987). Under inflationary expansion,
the energy density $\rho$ is positive and constant in time, but the
pressure is negative $p = -\rho$. This is said to be the `false
vacuum' state. Now, one of the most intriguing possibilities opened
up by inflation, is the possible creation of a universe `in a
laboratory'. Creation in a laboratory is taken to mean the creation
of a physical universe, by design, using the `artificial' means
available to an intelligent species. It is the ability of inflation
to maintain a constant energy density, in combination with a period
of exponential expansion, which is the key to these laboratory
creation scenarios. The idea is to use a small amount of matter in
the laboratory, and induce it to undergo inflation until its volume
is comparable to that of our own observable universe. The energy
density of the inflating region remains constant, and because it
becomes the energy density of a huge region, the inflating region
acquires a huge total (non-gravitational) energy.

The original proposals for universe creation in a laboratory, made in
the late 1980s, suggested the use of false vacuum `bubbles'. A more
recent approach suggests the use of magnetic monopoles. Let us
consider each approach in turn.

In Farhi and Guth (1987), the creation of an inflationary universe in
the laboratory was considered to be a special case of the behaviour
of false vacuum bubbles. A false vacuum bubble is a region of such
vacuum surrounded by true vacuum. The study of false vacuum bubbles
made by Blau \textit{et al} (1987) forms the basis of the Farhi-Guth
proposal. In these models, they consider, for simplicity, a false
vacuum bubble to occupy, spatially, the interior of a solid ball
$\mathbb{D}^3$. In space-time, such a false vacuum bubble would
occupy the interior of a solid hyper-cylinder $\mathbb{R}^1 \times
\mathbb{D}^3$. Blau \textit{et al} consider a false vacuum bubble to
be surrounded by spherically symmetric, zero energy density true
vacuum. A `thin wall' separates the false vacuum from the infinite
region of true vacuum. A false vacuum bubble of radius above a
certain critical value will undergo inflation. The assumption that
the exterior region is spherically symmetric and empty, entails that
it must be a portion of the maximally extended Schwarzschild-Kruskal
black hole space-time. From this, Farhi and Guth infer that ``the
creation of a universe is necessarily associated with the production
of a black hole," (Farhi and Guth, 1987, p150). According to Farhi
and Guth, the created universe resides inside the event horizon of a
black hole, but appears to be a region of inflating false vacuum from
the inside.

\begin{figure}[h]
\centering
\includegraphics[scale=0.5]{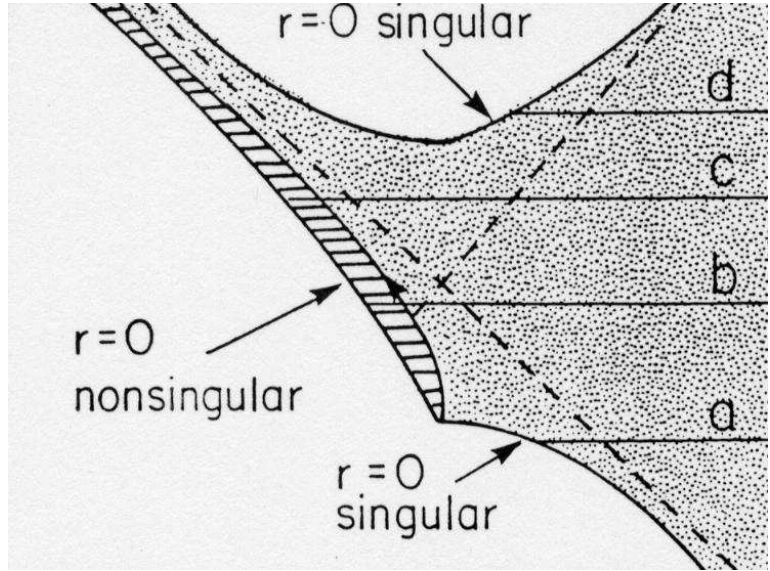}
\caption{Space-time diagram of a false vacuum bubble inside a
Schwarzschild black hole, from Guth (1991), p239.} \label{Bubble1}
\end{figure}

The Schwarzschild-Kruskal space-time has topology $\mathbb{R}^2
\times S^2$. Diagrammatically, it is often represented by the Kruskal
diagram, in which the spherical dimensions are suppressed, and only
the geometry on $\mathbb{R}^2$ is delineated. The paper of Blau
\textit{et al} provides the means to smoothly join an inflating
region of false vacuum to a portion of Schwarzschild-Kruskal
space-time. The interface between the two regions is described by a
curve in the Kruskal diagram. Each point of the Kruskal diagram
corresponds to a 2-sphere in the 4-dimensional space-time, hence a
curve in the Kruskal diagram corresponds to a hyper-cylinder
$\mathbb{R}^1 \times S^2$ in the 4-dimensional space-time. As
depicted in Figure \ref{Bubble1}, with the Kruskal diagram oriented
so that the null curves lie at $45\deg$ to the vertical, one removes
from the diagram the region which lies to the left of the
interface-curve, and one attaches the interior of the inflating false
vacuum bubble in its place. The inflating false vacuum bubble has the
space-time topology $\mathbb{R}^1 \times \mathbb{D}^3$, and one joins
the $\mathbb{R}^1 \times S^2$ boundary of the bubble to the
$\mathbb{R}^1 \times S^2$ boundary of the remaining
Schwarzschild-Kruskal space-time.

Blau \textit{et al} then introduce a preferential foliation of the
resulting space-time. This foliation is depicted as a horizontal
slicing of the doctored Kruskal diagram. The corresponding
foliation of the undoctored Kruskal diagram yields the
Einstein-Rosen bridge, a non-traversable wormhole.

\begin{figure}[h]
\centering
\includegraphics[scale=0.5]{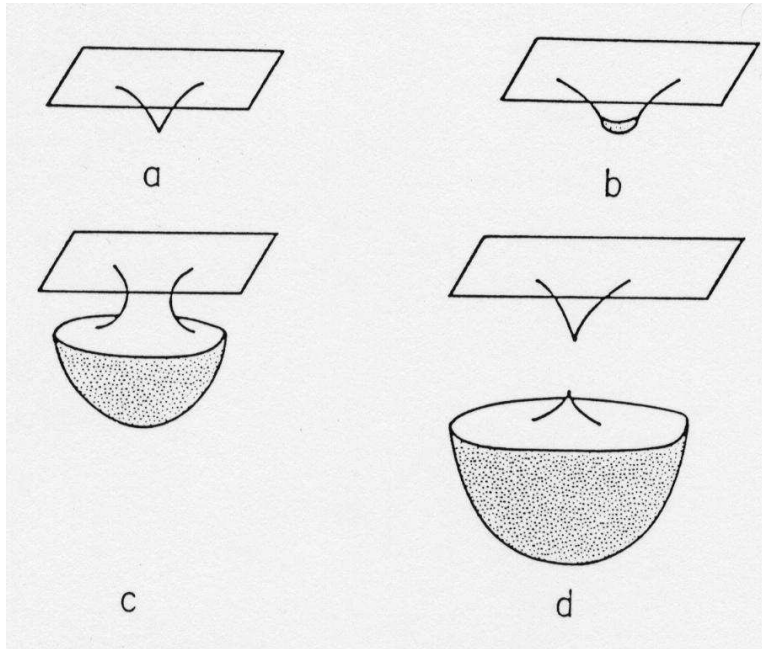}
\caption{False vacuum bubble detaching from its parent, from Guth
(1991), p240.} \label{Bubble2}
\end{figure}

Running through the leafs of the foliation, as depicted in Figures
\ref{Bubble1} and \ref{Bubble2}, one begins with a region of space
which is asymptotically flat at large distances, but which contains a
gravitational `sink' inside the Schwarzschild radius at $r = 2M$. A
finite region of false vacuum then appears inside the Schwarzschild
radius, and in successive leafs of the foliation, it mushrooms in
size. The bubble grows, however, on what would have been the white
hole side of the Einstein-Rosen bridge. As Guth puts it ``the
swelling takes place by the production of new space; the plane of the
original space is unaffected." (Guth 1991, p239-240). Running through
the foliation, there is a similar `pinching off' effect to that
encountered by the Einstein-Rosen bridge. The inflating bubble
separates from the rest of space like a rain drop which hangs from
the gable of a window, elongates, and then detaches itself. Under
this particular foliation, the spatial slices become disconnected,
one component consisting of the inflating bubble, the other
consisting of the throat of a black hole and the surrounding,
asymptotically flat space. An expanding false vacuum child universe
is spatially connected to a parent universe for a short time, before
the umbilical cord is severed, and the parent becomes spatially
disconnected from the child. Guth states that the false vacuum bubble
``completely disconnects from the original space-time, forming a new,
isolated closed universe." (Guth 1991, p240). Although the child
universe does indeed become spatially disconnected from the parent
universe after a period of time, the \emph{space-time} is very much
connected; it is merely the spacelike hypersurfaces along one
particular foliation which become disconnected. Because the
Einstein-Rosen bridge is a non-traversable wormhole, once the child
universe has begun to inflate, it is causally disconnected from the
parent universe. The inflating bubble of false vacuum resides in the
white hole part of the Kruskal diagram, and there is no timelike or
null curve from the black hole region to the white hole region.

One might infer that the creation of a child universe is confined to
the interior of a black hole to prevent the laboratory from being
engulfed by the expansion of the new universe. This, however, is not
correct, for even if an inflating false vacuum bubble were to be
created outside a black hole, it would not expand to engulf the
creator and his laboratory. The region of false vacuum would have
negative pressure, and would therefore be at lower pressure than its
surroundings. As Guth states, for an observer in the exterior region,
``the pressure gradient would point inward and the observer would not
expect to see the region increase in size." (Guth 1991, p238).

Farhi and Guth are not troubled by the prospect of a child universe
created inside the Schwarzschild radius of a black hole. They state
that it ``does not in principle present an insurmountable obstacle"
(Farhi and Guth, p150) to the creation of a universe by man-made
processes. They argue that ``ordinary materials (e.g. stars) can
collapse to form black holes, and it is possible at least to conceive
of a laboratory setup that would produce a black hole," (ibid.).
However, they point out that on a classical level, the creation of an
inflationary universe in the laboratory requires the presence of an
initial singularity. This is a serious obstacle because, as Guth puts
it, ``although an initial singularity is often hypothesized to have
been present at the big bang, there do not appear to be any
singularities available today." (Guth 1991, p240).

In the doctored Kruskal diagram, it is the white hole singularity
which is present in the early slices of the foliation, and which
provides the initial singularity. The final singularity of the
doctored Kruskal diagram is not an impediment as it is something
which could be created as a consequence of the universe-creation
process. The difficulty with an initial singularity is that it
would be a necessary precursor to the creation of a child
universe.

The presence of an initial singularity does not follow from the
supposition that child universe creation takes place within a
black hole/white hole spacetime; it follows from modelling the
false vacuum bubble as a part of de Sitter space-time. Farhi and
Guth use a reverse version of Penrose's 1965 singularity theorem
to establish their claim; this version deals with past trapped
surfaces rather than the future trapped surfaces which are the
hallmark of gravitational collapse. Farhi and Guth refer to the
past-trapped surfaces as `anti-trapped' surfaces. A sufficiently
large false vacuum bubble will contain anti-trapped two-spheres.
Farhi and Guth use the theorem:

A space-time $(\mathcal{M},g)$ contains an initial singularity if

\begin{enumerate}
\item{The null-convergence condition is satisfied: $Ric(v,v) \geq 0$
for all null vectors $v$.}
\item{There exists a non-compact Cauchy
hypersurface in $(\mathcal{M},g)$.}
\item{There exists an
anti-trapped compact surface in $(\mathcal{M},g)$.}
\end{enumerate}

Farhi and Guth (p154) argue that prior to the compression required
for universe creation, the laboratory space-time would be closely
approximated by Minkowski space-time. They infer from this that
condition 2) would be satisfied. Condition 1) is equivalent to the
`very weak' energy condition, $T(v,v) \geq 0$ for any null vector
$v$. If the weak energy condition is satisfied, it entails that
the very weak energy condition must be satisfied. Farhi and Guth
point out that the weak energy condition, hence the very weak
energy condition, is satisfied by all standard classical models of
matter fields. They show that in the spherically symmetric
idealization, condition 3) is satisfied, hence there must be an
initial singularity. They also provide good reasons for believing
that even a lack of spherical symmetry cannot avoid anti-trapped
surfaces and an initial singularity.

It must be emphasised again that the initial singularity is not a
consequence of the assumption that a man-made universe must be
embedded inside a black hole space-time. In the doctored Kruskal
diagram, the two-spheres represented by points inside the white hole
region are indeed past-trapped (`anti-trapped'), but it is the
past-trapped two-spheres of the inflating false-vacuum bubble which
entail the initial singularity. An inflating false vacuum bubble
which is not surrounded by $Ric = 0$ true vacuum, but by the more
realistic matter fields of a laboratory, would still require an
initial singularity.

The belief that the creation of a child universe cannot be achieved
classically, prompted the suggestion that it could be achieved by
quantum tunnelling instead. It was suggested that a false vacuum
bubble of radius below the critical value for inflation, but free
from an initial singularity, might be able to quantum mechanically
tunnel into an inflationary state (see Ansoldi and Guendelman 2006,
for details and references). More generally, certain quantum effects
seem to suggest that the energy conditions of general relativity can
be violated. Farhi and Guth (p154) allude to the fact that quantum
field theory in curved space-time generally violates the very weak
energy condition, and this implies that the creation of a child
universe could occur without an initial singularity.

Whilst the original idea for universe creation in a laboratory
proposed a false-vacuum bubble, represented as part of de Sitter
space-time, embedded inside the maximally extended
Schwarzschild-Kruskal spacetime, Sakai \textit{et al} (2006) imagine
a magnetic monopole, also represented as part of de Sitter
space-time, but embedded inside maximally extended
Reissner-Nordstr\"om space-time. To be more precise, a part of de
Sitter space-time is joined to a part of the maximally extended
Reissner-Nordstr\"om space-time for the case where the mass exceeds
the charge, $q<M$. Sakai \textit{et al} show that a classically
stable monopole could evolve into an inflationary universe by a
classical process, without quantum tunnelling. If the mass of the
monopole exceeds its charge, then it becomes inflationary. Sakai
\textit{et al} propose that the accretion or implosion of mass onto
an initially stable monopole, be represented by a spherical `domain
wall', surrounding the monopole, which eventually collides with it.

\begin{figure}[h]
\centering
\includegraphics[scale=0.5]{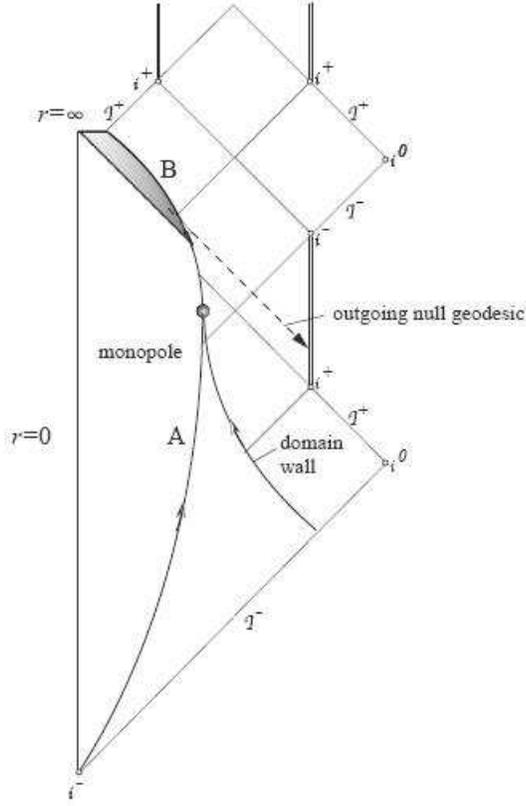}
\caption{Diagram of an inflating monopole inside a
Reissner-Nordstr\"om space-time, from Sakai \textit{et al} 2006.}
\label{Monopole}
\end{figure}

The significance of the maximally extended Reissner-Nordstr\"om
space-time for $q<M$ is that it includes a timelike singularity which
does not belong to the past of the region outside the black hole
containing the laboratory. The Reissner-Nordstr\"om space-time is
specified by the following metric tensor on $\mathbb{R}^2 \times
S^2$:

$$
ds^2 = -\left(1-\frac{2M}{r}+\frac{q^2}{r^2}\right)dt \otimes dt +
\left(1-\frac{2M}{r}+\frac{q^2}{r^2}\right)^{-1} dr \otimes dr +
r^2 d\Omega^2 \;.
$$ $d\Omega^2$, of course, is the standard metric on the 2-sphere. Unlike the
Kerr space-time (for a rotating black hole), the Reissner-Nordstr\"om space-time
doesn't have a ring singularity, and cannot be extended to
negative values of $r$. However, just like the Kerr space-time, it
possesses an outer horizon $r_+$, and an inner horizon $r_-$:

$$
r_\pm = M \pm (M^2 - q^2)^{1/2} \;.
$$ These horizons are defined by the roots of the function

$$
\Delta = r^2 - 2Mr + q^2 \;,
$$ i.e., the values of $r$ at which $\Delta$ is zero. To obtain the maximally extended
space-time, one first partitions the Reissner-Nordstr\"om geometry
into three distinct blocks: (i) the asymptotically flat block outside
the black hole, with $r \in [r_+,\infty)$; (ii) the inter-horizon
block, with $r \in [r_-, r_+]$; and (iii) the black hole interior,
with $r \in [r_-,0)$, and the singularity `at' $r=0$. The maximally
extended space-time for $q< M$ is obtained by tessellating various
copies of these blocks into an infinite chain.

The created inflationary universe depicted in Figure \ref{Monopole}
includes past incomplete null geodesics emanating from anti-trapped
surfaces, but there is no initial singularity as such. ``Although a
singularity exists in the past of the inflating monopole, the
singularity is located in the future of the experimenter in the
laboratory. In other words, even if no singularity exists in the past
of the experimenter who makes a monopole, inflation in the monopole
is realizable in the future of the experimenter. From a observational
point of view, however, since the inflating monopole is realized
inside a black hole, the experimenter cannot observe it unless he or
she enters into the black hole," (Sakai \textit{et al}, 2006). The
inflating monopole could be created by an experimenter whose past is
geodesically complete.

Magnetic monopoles are predicted to exist by certain unified field
theories, and whilst a magnetic monopole has yet to be discovered, a
collision between an electron and a positron could, in principle,
create a monopole--anti-monopole pair. Monopoles have masses much
greater than those of electrons and positrons, however, and the
kinetic energies required to create them by such a collision are
beyond the capabilities of contemporary particle accelerators.
Universe creation in a laboratory therefore remains beyond current
technology, but theoretically possible.


\begin{thebibliography}{99}
\bibitem{Ansoldi06}
Ansoldi, S., Guendelman, E.I. (2006). Child Universes in the
Laboratory. arXiv:gr-qc/0611034.
\bibitem{Blau87}
Blau, S.K., Guendelman, E.I., Guth, A.H. (1987). Dynamics of False
Vacuum bubbles, \emph{Physical Review} D35, pp1747-1766.
\bibitem{BlauGuth87}
Blau, S.K., Guth, A.H. (1987). Inflationary Cosmology, in
\emph{300 Years of Gravitation}, S.W.Hawking and W.Israel (eds.),
pp524-603. Cambridge: Cambridge University Press.
\bibitem{Derdzinski92}
Derdzinski, A. (1992). \emph{Geometry of the Standard Model of
Elementary Particles}, Texts and Monographs in Physics,
Berlin-Heidelberg-New York: Springer Verlag.
\bibitem{FarhiGuth87}
Farhi, E., Guth, A.H. (1987). An obstacle to creating a universe
in the laboratory, \emph{Physics Letters} B183(2), pp149-155.
\bibitem{Farhi90}
Farhi, E., Guth, A.H., Guven, J. (1990). Is it Possible to Create a
Universe in the Laboratory by Quantum Tunneling, \emph{Nuclear
Physics} B339, pp417-490.
\bibitem{Guth91}
Guth, A.H. (1991). Can a Man-Made Universe be Created by Quantum
Tunneling without an Initial Singularity?, \emph{Physica Scripta} Vol
T36, pp237-246.
\bibitem{Sakai06}
Sakai, N., Nakao, K., Ishihara, H., and Kobayashi, M. (2006). Is it
possible to create a universe out of a monopole in the laboratory?
\emph{Physical Review} D74(2), 024026.
\bibitem{Tegmark05}
Tegmark, M., Aguirre, A., Rees, M.J., and Wilczek, F. (2005).
Dimensionless constants, cosmology and other dark matters,
arXiv:astro-ph/0511774.
\end{thebibliography}
\end{document}